\AtBeginDocument{%
  }
\documentclass[sigconf]{acmart}

\copyrightyear{2026}
\acmYear{2026}
\setcopyright{cc}
\setcctype{by}
\acmConference[MM '26] {Proceedings of the 34th ACM International Conference on Multimedia}{November 10--14, 2026}{Rio de Janeiro, Brazil.}
\acmBooktitle{Proceedings of the 34th ACM International Conference on Multimedia (MM '26), November 10--14, 2026, Rio de Janeiro, Brazil}
\acmISBN{979-8-4007-2213-4/2026/11}
\acmDOI{10.1145/3767308.3836120}

\usepackage{microtype}
\usepackage{graphicx}
\usepackage{booktabs}
\usepackage{multirow}
\usepackage{tcolorbox}
\tcbuselibrary{breakable}
\usepackage{xcolor}
\usepackage{pifont}
\usepackage{float}
\usepackage{amsmath}
\usepackage{mathtools}
\usepackage{amsthm}
\usepackage[capitalize,noabbrev]{cleveref}

\theoremstyle{plain}

\theoremstyle{definition}

\theoremstyle{remark}

\definecolor{osgreen}{RGB}{235,245,238}
\definecolor{csviolet}{RGB}{244,238,250}
\definecolor{oursblue}{RGB}{235,242,250}

\newcommand{\YesIcon}{\textcolor{green!60!black}{\ding{51}}}
\newcommand{\rev}[1]{\textcolor{black}{#1}}
\newcommand{\NoIcon}{\textcolor{red!70!black}{\ding{55}}}
\newcommand{\best}[1]{\textbf{#1}}
\newcommand{\second}[1]{\underline{#1}}

\begin{document}

\title[OmniClimate-TC]{OmniClimate-TC: Physics-Aware Visual Abstractions for Multimedia Reasoning over Tropical Cyclones}

\author{Luwei Xiao}
\authornote{Both authors contributed equally to this research.}
\affiliation{%
  \institution{National University of Singapore}
  \city{Singapore}
  \country{Singapore}
}
\author{Xin Wang}
\authornotemark[1]
\affiliation{%
  \institution{National University of Singapore}
  \city{Singapore}
  \country{Singapore}
}

\author{Keane Ong}
\affiliation{%
  \institution{National University of Singapore}
  \city{Singapore}
  \country{Singapore}
}

\author{Jiawen Wei}
\affiliation{%
  \institution{National University of Singapore}
  \city{Singapore}
  \country{Singapore}
}

\author{Chenyu Dong}
\affiliation{%
  \institution{National University of Singapore}
  \city{Singapore}
  \country{Singapore}
}
\author{Rui Mao}
\affiliation{%
  \institution{Nanyang Technological University}
  \city{Singapore}
  \country{Singapore}
}

\author{Erik Cambria}
\affiliation{%
  \institution{Nanyang Technological University}
  \city{Singapore}
  \country{Singapore}
}

\author{Gianmarco Mengaldo}
\authornote{Corresponding author: mpegim@nus.edu.sg}
\affiliation{%
  \institution{National University of Singapore}
  \city{Singapore}
  \country{Singapore}
}

\renewcommand{\shortauthors}{Luwei Xiao et al.}

\begin{abstract}
Meteorological reanalysis encodes extreme weather through continuous, physically constrained fields, posing a fundamental challenge for vision–language models (VLMs) whose perceptual assumptions are shaped by natural images. 
Tropical cyclones exemplify this mismatch: critical properties such as intensity extrema, asymmetry, spatial extent, and physical impacts arise from field-level organization rather than object-centric visual cues. 
Existing approaches address this gap through text alignment or annotation, treating the problem as multimodal supervision rather than representation design.
We introduce Physics-Aware Visual Abstraction (PAVA), a plug-and-play physics-aware representation and annotation interface that maps physical reanalysis fields to visually identifiable and semantically grounded perceptual abstractions for supervision and evaluation in vision--language reasoning. 
Building on PAVA, we construct OmniClimate-TC, a benchmark for tropical cyclone analysis spanning five classes of reasoning and nine tasks, with 243,890 physically grounded instruction-tuning pairs. 
Using PAVA-aligned supervision, we adapt VLMs and provide evidence that this representation design improves reasoning over tropical cyclone hazard fields. 
Our results position OmniClimate-TC as a benchmark for multimedia reasoning over structured geophysical fields, and highlight representation design as a key ingredient for physically grounded reasoning in scientific media.
\end{abstract}



\maketitle

\section{Introduction}
\label{sec:intro}

Extreme weather events pose fundamental challenges for visual-language models (VLMs), because the relevant signals are encoded in high-dimensional numerical data representing physically constrained spatiotemporal fields~\cite{bergen2019machine, camps2025artificial}.
\rev{Among these phenomena, tropical cyclones are an important testbed: their evolution and hazard footprint arise from tightly coupled physical variables and coherent organization across scales~\cite{reichstein2019deep, dong2024indo}, that are captured by numerical datasets, such as meteorological reanalysis products (e.g., ERA5~\cite{hersbach2020era5}).
These numerical datasets are composed of continuous fields, and the semantics they encode differ fundamentally from the object-centric visual structure of natural images~\cite{koldasbayeva2024challenges}.}
\rev{The rapid growth of AI-based weather models further sharpens the need for automated semantic interpretation, as high-fidelity forecast fields increasingly exceed what can be analyzed consistently by human experts at scale~\cite{reichstein2019deep, camps2025artificial, dong2025time}.}

VLMs have recently shown strong multimodal perception and reasoning capabilities~\cite{zhang2024vision, steyvers2025large, yang2025visionzip}. \rev{However, their visual encoders and pretraining corpora are dominated by natural images, favoring object-centric semantics and image-native inductive biases~\cite{lee2024vhelm, zhu2025ibd, laurenccon2024matters}. Meteorological reanalysis does not conform to these assumptions, as it comprises continuous spatiotemporal fields in which meaning arises from extrema, gradients, coherence, asymmetry, and scale-dependent structure rather than from discrete objects~\cite{koldasbayeva2024challenges, studholme2022poleward, patrick2022general}. The result is a representational mismatch: the standard ``image'' interface does not make explicit the physically meaningful attributes to which language-based reasoning must refer to.}
Recent studies have attempted to bridge this gap by aligning reanalysis plots with text, including captioning~\cite{li2025cllmate}, question answering over climate maps~\cite{ma2024weatherqa}, report generation~\cite{tang2025meteorpred}, and LLM-assisted annotation~\cite{chen2025climateiqa}. While effective for producing multimodal supervision, these approaches implicitly assume that the relevant physical semantics are visually available and can be learned end-to-end from paired data. In practice, many core meteorological concepts (e.g., intensity extrema, asymmetry, and extent) are operator-defined over space and scale, not directly readable as ``objects'' in raw heatmaps. Consequently, annotations are expensive and difficult to standardize, and scaling supervision alone does not resolve the underlying perceptual gap. \rev{\textit{This suggests that the central challenge is representation design: constructing physics-consistent perceptual abstractions that render field semantics visually identifiable and language-addressable.}}

\rev{Seen through a multimedia-reasoning lens, the problem is not just multimodal alignment; it is how to expose the right perceptual structure from scientific fields so that downstream reasoning remains grounded in the underlying physics.}
Guided by this diagnosis, we formalize our approach as Physics-Aware Visual Abstraction (PAVA), a plug-and-play paradigm that reframes extreme-weather understanding as a \textit{representation design} problem rather than a data-alignment problem. Instead of introducing an auxiliary inference-time modality, PAVA serves as a physics-aware representation and annotation interface that exposes semantically meaningful structure from meteorological reanalysis for benchmark construction, supervision design, and controlled evaluation. In the context of tropical cyclones, we instantiate PAVA by transforming continuous wind, precipitation, and sea-level pressure fields into cyclone-centric, visually identifiable abstractions that encode where extremes occur, how intensity is organized, how structure departs from symmetry, and how hazards extend over space. These abstractions constitute a shared perceptual interface that makes physically grounded concepts accessible to multimodal reasoning and enables systematic construction of supervision across complementary reasoning dimensions. Building on this framework, we introduce OmniClimate-TC, a novel benchmark that spans recognition and impact-level reasoning over tropical cyclones, and use it to validate ClimateTCX, a two-stage adaptation of vision–language models that progressively composes perceptual grounding into physical impact analysis. 
In a ntushell, our contributions are as follows:
\begin{itemize}
    \item \rev{We introduce \textbf{Physics-Aware Visual Abstraction (PAVA)}, a physics-grounded representation/annotation interface that maps continuous reanalysis fields to visually identifiable and semantically grounded perceptual abstractions for supervision and evaluation in vision--language reasoning.} 
    \item \rev{We construct \textbf{OmniClimate-TC}, a tropical-cyclone benchmark for multimedia reasoning that spans five categories of visual perception and compositional reasoning, covering nine tasks and comprising 243,890 physically grounded instruction-tuning pairs.}
    \item \rev{We develop \textbf{ClimateTCX}, a family of vision--language models adapted to OmniClimate-TC for physics-grounded multimedia reasoning.}
    \rev{Across multiple tasks, ClimateTCX consistently outperforms strong baselines, supporting the utility of PAVA-aligned representation design for physics-grounded reasoning over hazard fields.}
\end{itemize}
\begin{table*}[t!]
\setlength\tabcolsep{2pt}
\centering
\caption{
Comparison to related Earth-system multimodal benchmarks relevant to multimedia reasoning.
\textbf{Physics-grounded label} indicates whether labels/questions are derived from physical continuous fields, rather than post-hoc textual annotation.
}
\label{tab_climate_data}
\resizebox{0.95\textwidth}{!}{
\begin{tabular}{llccccccc}
\toprule
\textbf{Dataset} &
\textbf{Event type} &
\textbf{Variables} &
\textbf{Volume} &
\textbf{Coverage} &
\textbf{Time span} &
\textbf{Physics-grounded} &
\textbf{Physical impact} \\
 & & & & & & \textbf{labels} & \textbf{reasoning included?} \\
\midrule

Mesogeos~\cite{kondylatos2023mesogeos}
& Wildfire / Earth system
& Reanalysis
& 25,722
& Regional
& 2006--2022
& \NoIcon
& \NoIcon \\

WeatherQA~\cite{ma2024weatherqa}
& Severe convection
& Reanalysis, Obs.
& 8,511
& Regional
& 2014--2020
& \NoIcon
& \NoIcon \\

RadarQA~\cite{heradarqa}
& Radar-based weather
& Reanalysis
& 70,000
& Worldwide
& N/A
& \NoIcon
& \NoIcon \\

ClimateIQA~\cite{chen2025climateiqa}
& General weather
& Reanalysis
& 762,120
& Worldwide
& 2023
& \NoIcon
& \NoIcon \\

CLLMate~\cite{li2025cllmate}
& Weather events
& Reanalysis
& 7,747
& Worldwide
& 2015--2023
& \NoIcon
& \NoIcon \\

OmniEarth-Bench~\cite{wang2025omniearth}
& Multi-sphere Earth obs.
& Remote sensing
& 29,855
& Worldwide
& 2010--2025
& \NoIcon
& \NoIcon \\

MP-Bench~\cite{tang2025meteorpred}
& Severe weather
& Reanalysis
& 421,363
& Regional
& 2023--2024
& \NoIcon
& \NoIcon \\

\midrule
\textbf{OmniClimate-TC}
& \textbf{Tropical cyclone}
& Reanalysis
& 243,890
& Worldwide
& 1979--2025
& \YesIcon
& \YesIcon \\

\bottomrule
\end{tabular}
}
\end{table*}

\section{Related Work}

\textbf{Vision-Language Models for Weather and Climate.} The intersection of meteorology and multimodal learning has seen rapid growth, driven by the need to interpret complex geoscientific data, such as reanalysis products. 
Early benchmarks like WeatherQA~\cite{ma2024weatherqa} and CLLMate~\cite{li2025cllmate} established text-field alignment, while OmniEarth-Bench~\cite{wang2025omniearth} broadened the scope to comprehensive Earth systems. Recent specialized works like ClimateIQA~\cite{chen2025climateiqa} and MeteorPred~\cite{tang2025meteorpred} have further integrated contour tracking and 4D spatiotemporal data.
While these contributions have established a strong foundation for meteorological VLMs, they predominantly adopt an \textbf{implicit visual learning} paradigm, relying on the model to infer meteorological semantics directly from heatmap textures via end-to-end supervision.
However, meteorological fields differ fundamentally from natural images, as their interpretability relies on rigorous physical constraints (e.g., conservation laws) rather than object-centric visual cues. 
\textbf{PAVA} seeks to bridge this gap by introducing \textbf{explicit physical abstractions}, that is: transforming continuous fields into discrete, physically grounded attributes that provide a structured interface between raw physical signals and the reasoning capabilities of vision--language models. As shown in Tab.~\ref{tab_climate_data}, OmniClimate-TC, built on PAVA, offers a complementary benchmark with explicit physics-grounded supervision and physical-impact reasoning.

\textbf{Extreme Weather Hazards and the Semantic Interpretation Gap.}
In weather forecasting, the field has rapidly evolved from early baselines (e.g., Pangu-Weather~\cite{bi2022pangu}, GraphCast~\cite{lam2023learning}) to advanced architectures like GenCast~\cite{price2025probabilistic} and Aurora~\cite{bodnar2025foundation}. AI approaches in climate modeling have advanced from pioneering hybrid modeling like NNCAM~\cite{wang2022stable} and differentiable NeuralGCM~\cite{kochkov2024neural} to recent frameworks like CondensNet~\cite{wang2026condensnet} and ACE2~\cite{wattmeyer2025ace2} that enforce physical constraints for stable multi-decadal simulation.
However, the raw outputs produced by these and more traditional models lack explicit hazard \textit{semantics}, creating a critical dependency on manual expert interpretation that is both unscalable and error-prone. 
As data volume explodes, the widening gap between information availability and human cognitive capacity~\cite{reichstein2019deep, camps2025artificial} risks obscuring vital warning signals. 
In this work, we mitigate this bottleneck by proposing \textbf{PAVA}, a physics-aware interface that distills high-dimensional numerical fields into structured hazard representations, supporting more systematic hazard-oriented interpretation.

\begin{figure*}[t]
    \centering
    \includegraphics[width=0.9\linewidth]{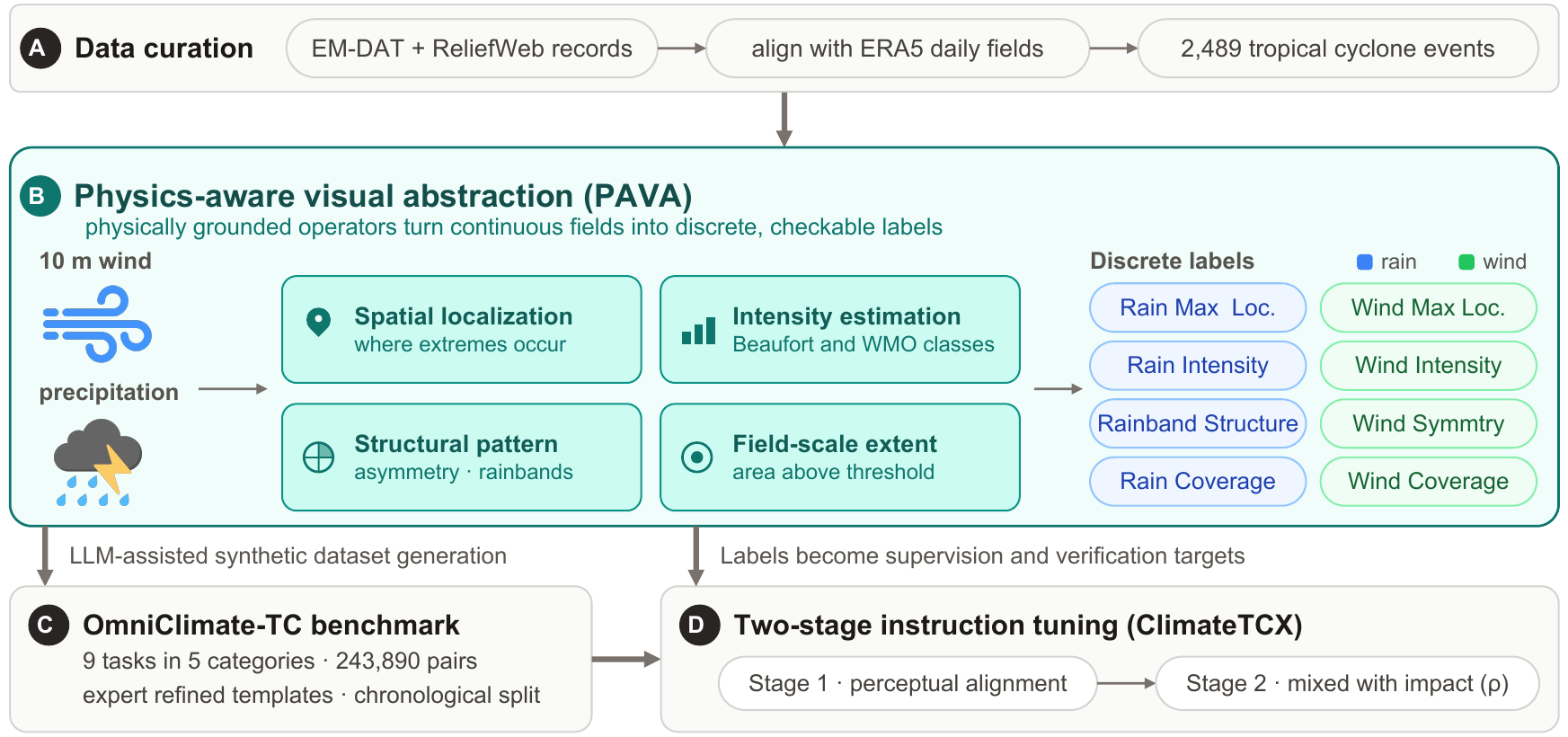}
    \caption{Overall architecture of the proposed framework. This pipeline consists of four components: Dataset Collection \& Curation, Physics-Aware Visual Abstraction (PAVA), OmniClimate-TC Construction and Two-stage Instruction-tuning.}
    \label{fig:climate_pava}
\end{figure*}

\begin{figure*}[t]
    \centering
    \includegraphics[width=0.9\linewidth]{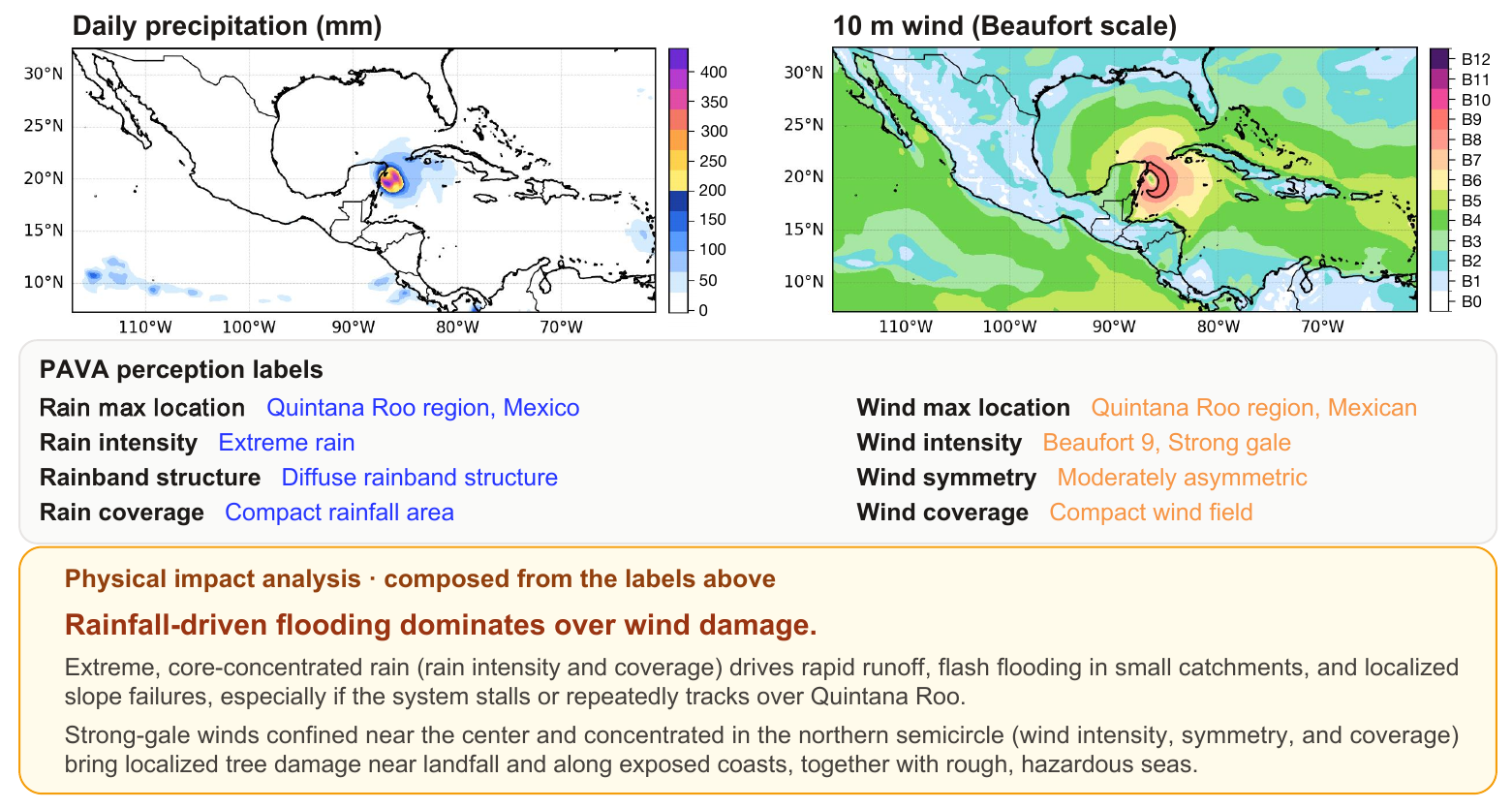}
    \caption{An example from OmniClimate-TC.
\textcolor{blue}{Blue} and \textcolor{orange}{Orange} denote PAVA-derived labels from the precipitation field
and the wind field (e.g., spatial location, intensity, structure, and coverage), respectively.
}

    \label{adix:example}
\end{figure*}

\section{Method}
\label{sec:method}

\textbf{Problem Setup and Motivation.}
For a tropical cyclone event, the input consists of spatiotemporal physical fields:
\begin{equation}
\small
\mathcal{F} = \{ F_t^{(k)}(\phi,\lambda) \mid k = 1,\dots,K,\; t \in \mathcal{T} \},
\label{eq:input_fields}
\end{equation}
where $(\phi,\lambda)$ denote latitude and longitude, and $F_t^{(k)}$ is the $k$-th physical variable (e.g., wind, precipitation, or sea-level pressure) at time $t$. In this work, $\mathcal{F}$ is derived from daily ERA5 reanalysis over the event duration $\mathcal{T}$.
Unlike object-centric visual inputs, reanalysis fields encode event semantics through spatial organization and physically constrained patterns rather than discrete entities, creating a fundamental mismatch with multimodal perceptual interfaces and motivating physics-aligned visual abstractions. The overall framework of our proposed method is shown in Fig.~\ref{fig:climate_pava}.

\subsection{Dataset Collection \& Curation}
\label{sec:data_collection}

We construct an event-centric tropical cyclone corpus by aligning historical disaster
records with global meteorological reanalysis.
Meteorological fields are obtained from ERA5 daily single-level reanalysis~\cite{hersbach2020era5}.
We opt for variables that are directly relevant
to TC hazards and consistently available worldwide: 10\,m wind components
($u_{10}$, $v_{10}$), daily total precipitation ($tp$), and mean sea-level pressure ($msl$).
The dataset covers tropical cyclone events from January 1979 to October 2025.
Event records are compiled from EM-DAT and ReliefWeb, merged across sources to remove
duplicates and overlaps, yielding 2,489 unique TC events.
Each event is associated with a spatiotemporal window and a geographic subregion of interest.
Since disaster records typically report affected countries rather than precise storm
coordinates, we infer event-level locations via a dedicated geocoding procedure:
subnational locations are resolved using standardized gazetteers\footnote{\url{https://www.geonames.org/}}, and unresolved cases
are assigned country-level centroids.

\subsection{Physics-Aware Visual Abstraction (PAVA) }

To bridge the representation gap mentioned above, we propose physics-aware visual abstractions (PAVA) that maps continuous reanalysis fields to discrete, interpretable labels. We first localize a tropical cyclone by combining the subregional event-level locations with sea-level pressure fields.

\textbf{Spatial Localization.}
Spatial localization aligns visually salient regions in hazard maps with physically dominant locations. Let $\Omega_e$ denote the event-specific spatial domain (i.e., the set of spatial grid points within the event-centric subregion bounding box).
Following the definition in Eq.~\ref{eq:input_fields}, let $P_t$ and $W_t$ denote the instantaneous precipitation and wind fields in $\mathcal{F}$. We construct the \textbf{event-aggregated} hazard field $F_e$ for a physical variable $F \in \{P, W\}$ by taking the temporal maximum over the duration $\mathcal{T}$:
\begin{equation}
\small
F_e(\phi,\lambda) = \max_{t \in \mathcal{T}} F_t(\phi,\lambda).
\end{equation}
The spatial location of the hazard extremum is identified as:
\begin{equation}
\small
(\phi_f,\lambda_f) = \arg\max_{(\phi,\lambda)\in\Omega_e} F_e(\phi,\lambda).
\end{equation}
Instantiating $F$ with $P$ and $W$ yields the peak coordinates $(\phi_r,\lambda_r)$ and $(\phi_w,\lambda_w)$, respectively.
The labels \texttt{Rain-Max-Loc} and \texttt{Wind-Max-Loc} provide spatial anchors that tie visual peaks in heatmaps to physically meaningful extrema. \rev{These coordinates are then mapped, via the Google Geocoding API, to standardized country and first-level administrative units, yielding multi-scale evaluation targets rather than free-form place names.}

\textbf{Intensity Estimation.}
Intensity estimation maps physical magnitude to visually interpretable severity.
We compute event-level intensity maxima:
\begin{equation}
\small
P_{\max} = \max_{(\phi,\lambda)\in\Omega_e} P_e(\phi,\lambda),
W_{\max} = \max_{(\phi,\lambda)\in\Omega_e} W_e(\phi,\lambda).
\end{equation}
These maxima are discretized into standard intensity categories: wind strength follows the Beaufort scale~\cite{huler2007defining}, and rainfall intensity uses physics-informed thresholds consistent with WMO-style daily extreme precipitation classifications\cite{clima2023guidelines}.
\rev{We formulate intensity estimation as classification since the benchmark is designed to recover meaningful severity categories rather than raw continuous magnitudes.}
This yields the \texttt{Max-Wind} and \texttt{Max-Rain} labels.

\textbf{Structural Pattern.}
Structural patterns capture mesoscale organization that is visually evident yet not reducible to pointwise values, such as asymmetry and banded precipitation structure. These patterns reflect dynamical processes of tropical cyclones.

\textit{Wind Symmetry.}
Using the wind-maximum location $(\phi_w,\lambda_w)$ as a reference, inspired by~\cite{klotz2017examination}, we partition $\Omega_e$ into four quadrants $\mathcal{Q}\in\{\mathrm{NE},\mathrm{NW},\mathrm{SE},\mathrm{SW}\}$. For each quadrant, we compute:
\begin{equation}
\small
S_{\mathcal{Q}}^{(w)} = \sum_{(\phi,\lambda)\in\mathcal{Q}} W_e(\phi,\lambda).
\end{equation}
We quantify the visual imbalance of the wind field using a normalized quadrant-contrast index:
\begin{equation}
\small
\alpha_w =
\frac{\max_{\mathcal{Q}} S_{\mathcal{Q}}^{(w)} - \min_{\mathcal{Q}} S_{\mathcal{Q}}^{(w)}}
{\sum_{\mathcal{Q}} S_{\mathcal{Q}}^{(w)} + \varepsilon},
\end{equation}
where $\varepsilon$ is a small constant for numerical stability (avoid division by zero).
In implementation, we discretize $\alpha_w$ as \texttt{symmetric} ($\alpha_w \le 0.15$), \texttt{moderately asymmetric} ($0.15 < \alpha_w \le 0.35$), and \texttt{strongly asymmetric} ($\alpha_w > 0.35$) to produce the \texttt{Wind-Sym} label.

\textit{Rainband Structure.}
Rainband structure characterizes whether precipitation is organized into
multiple coherent clusters or remains spatially diffuse.
Given the event-aggregated precipitation field $P_e(\phi,\lambda)$, we normalize it by:
\begin{equation}
\small
\tilde{P}_e(\phi,\lambda) =
\frac{P_e(\phi,\lambda)}{\max_{\Omega_e} P_e + \varepsilon},
\small
M_r(\phi,\lambda) =
\mathbf{1}\!\left[\tilde{P}_e(\phi,\lambda) \ge \tau_r \right],
\end{equation}
where $\tau_r$ is set to the 85th percentile of the normalized precipitation field, following common practice to isolate physically dominant rainfall structures\cite{schar2016percentile}. Connected-component analysis is applied to $M_r$ to identify spatially contiguous precipitation clusters~\cite{hoshen1976percolation}; in implementation, we assign \texttt{clustered} for $N_c \ge 3$, \texttt{weakly clustered} for $N_c = 2$, and \texttt{diffuse} otherwise.

\textbf{Field-scale Extent.}
Field-scale extent quantifies how broadly hazards are distributed in space, connecting physical footprint size to perceived spatial spread in visual representations.
Let $a(\phi)$ denote the physical area of a grid cell at latitude $\phi$, which scales as $a(\phi) \propto \cos(\phi)$ to correct for map projection distortions on the spherical Earth. Rain coverage~\cite{lonfat2004precipitation} is computed as:
\begin{equation}
\small
\mathrm{Cov}_r =
\frac{\sum_{i\in\Omega_e} \mathbf{1}\!\left[P_e(i) \ge \theta_r\right] \, a(\phi_i)}
{\sum_{i\in\Omega_e} a(\phi_i)},
\end{equation}
where $\theta_r = 50$ mm is a fixed heavy-rain threshold.
For wind, we adopt a relative threshold $\gamma$~\cite{niu2022assessing} tied to event intensity; we set $\gamma = 0.55$ and compute the wind coverage~\cite{holland1980analytic} as follows:
\begin{equation}
\small
\theta_w = \gamma W_{\max},
\small
\mathrm{Cov}_w =
\frac{\sum_{i\in\Omega_e} \mathbf{1}\!\left[W_e(i) \ge \theta_w\right] \, a(\phi_i)}
{\sum_{i\in\Omega_e} a(\phi_i)},
\end{equation}

The resulting labels \texttt{Rain-Cov} and \texttt{Wind-Cov} capture the spatial footprint of hazards, aligning physical coverage with visual extent in reanalysis maps.

\textbf{Unified Operator View of PAVA.}
We formalize PAVA as a family of physically grounded operators that map continuous reanalysis fields to discrete, visually interpretable representations.
Concretely, given an event-centric physical input as Eq.~\ref{eq:input_fields},
PAVA defines a set of abstraction operators:
\begin{equation}
\small
\mathcal{A} = \{ a_j \}_{j=1}^{J}, \qquad a_j:\; \mathcal{F}_e \rightarrow \mathcal{Y}_j,
\end{equation}
where each operator $a_j$ extracts a physically meaningful attribute $\mathcal{Y}_j$ from the underlying fields. By construction, each attribute is grounded in established meteorological diagnostics, discriminable across severity or structural categories, and expressed in standard terminology amenable to language-based reasoning. 

\subsection{OmniClimate-TC Construction}
Building on PAVA, we construct \textbf{OmniClimate-TC}, a structured benchmark designed to evaluate multimedia reasoning over tropical cyclone hazard fields.

\textbf{Event-centric visualization.}
For each record, 
Wind is visualized using a Beaufort-scale color mapping 
~\cite{huler2007defining}. Precipitation is visualized as a continuous field in physical units (mm), following common practice in reanalysis-based precipitation analysis~\cite{hersbach2020era5}.
All heatmaps are rendered within an event-centric geographic context defined by subregion based on the United Nations M49~\cite{united1982standard}.

\textbf{Instruction-tuning data.}
For each tropical cyclone event, we construct instruction--response pairs from templates constrained by PAVA-derived physical labels.
Templates are refined with domain experts, and GPT-5.1 is used only as a constrained text-realization backend after labels and templates are fixed.
\rev{For the first four perception categories, PAVA labels are deterministic supervision targets used only in benchmark construction and training, not auxiliary inference inputs. For Physical Impact Analysis, PAVA attributes constrain reference construction and verification.}
OmniClimate-TC spans five task categories covering nine tasks.
The first four categories correspond to \emph{visual perception} tasks that require direct interpretation of hazard-field patterns:
\textbf{Spatial Localization}, \textbf{Intensity Estimation},
\textbf{Structural Pattern}, and \textbf{Field-scale Extent}.
The fifth category, \textbf{Physical Impact Analysis}, is a \emph{compositional} reasoning task that synthesizes multiple physical cues into a concise, mechanism-focused hazard summary.

\begin{figure}[t]
    \centering
    \includegraphics[width=0.9\linewidth]{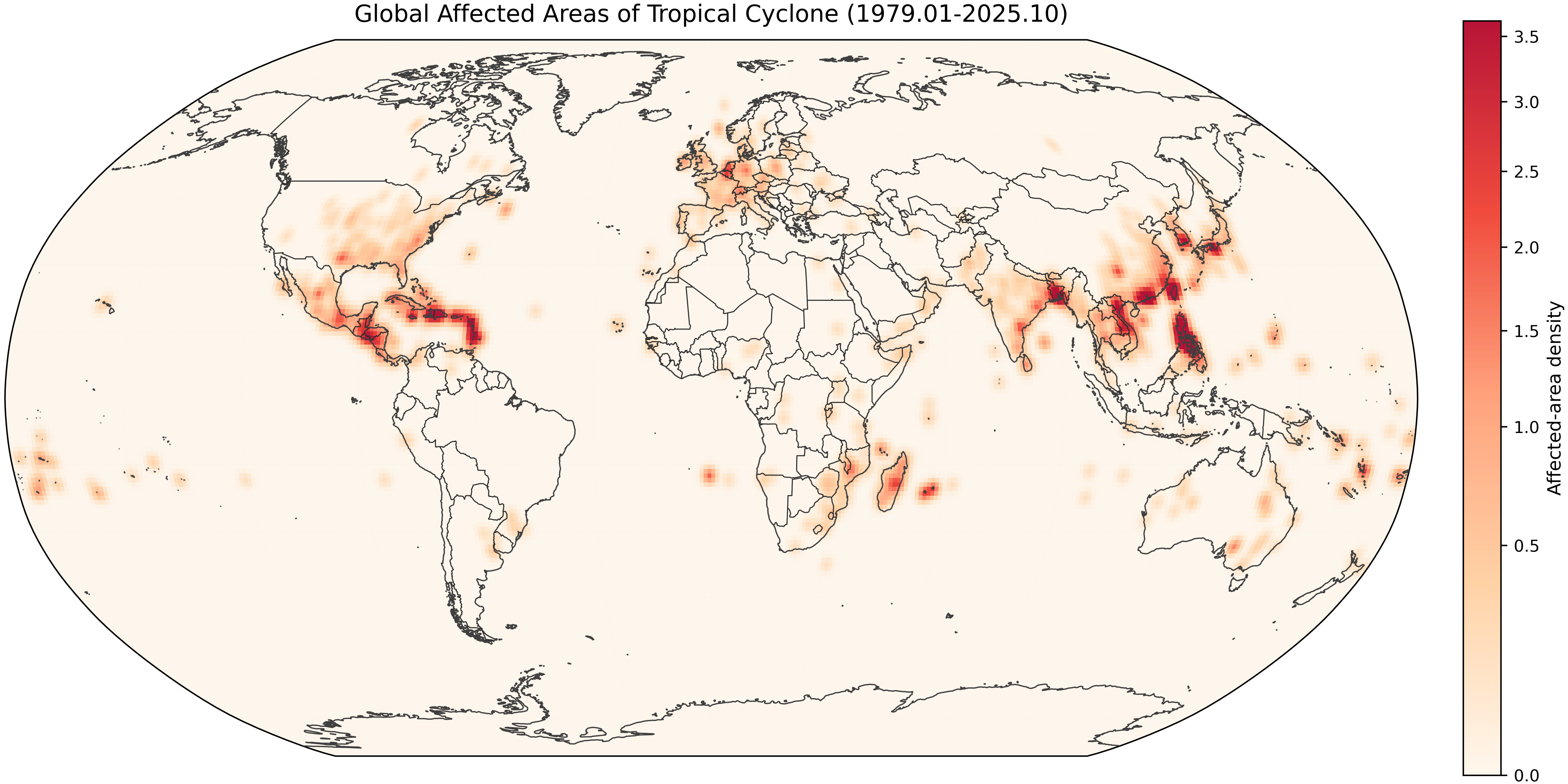}
    \caption{Density of tropical cyclone–affected areas from OmniClimate-TC,
showing broad coverage across the world.}
    \label{fig:climate_tc_map}
\end{figure}
\begin{figure}[t]
    \centering
    \includegraphics[width=0.8\linewidth]{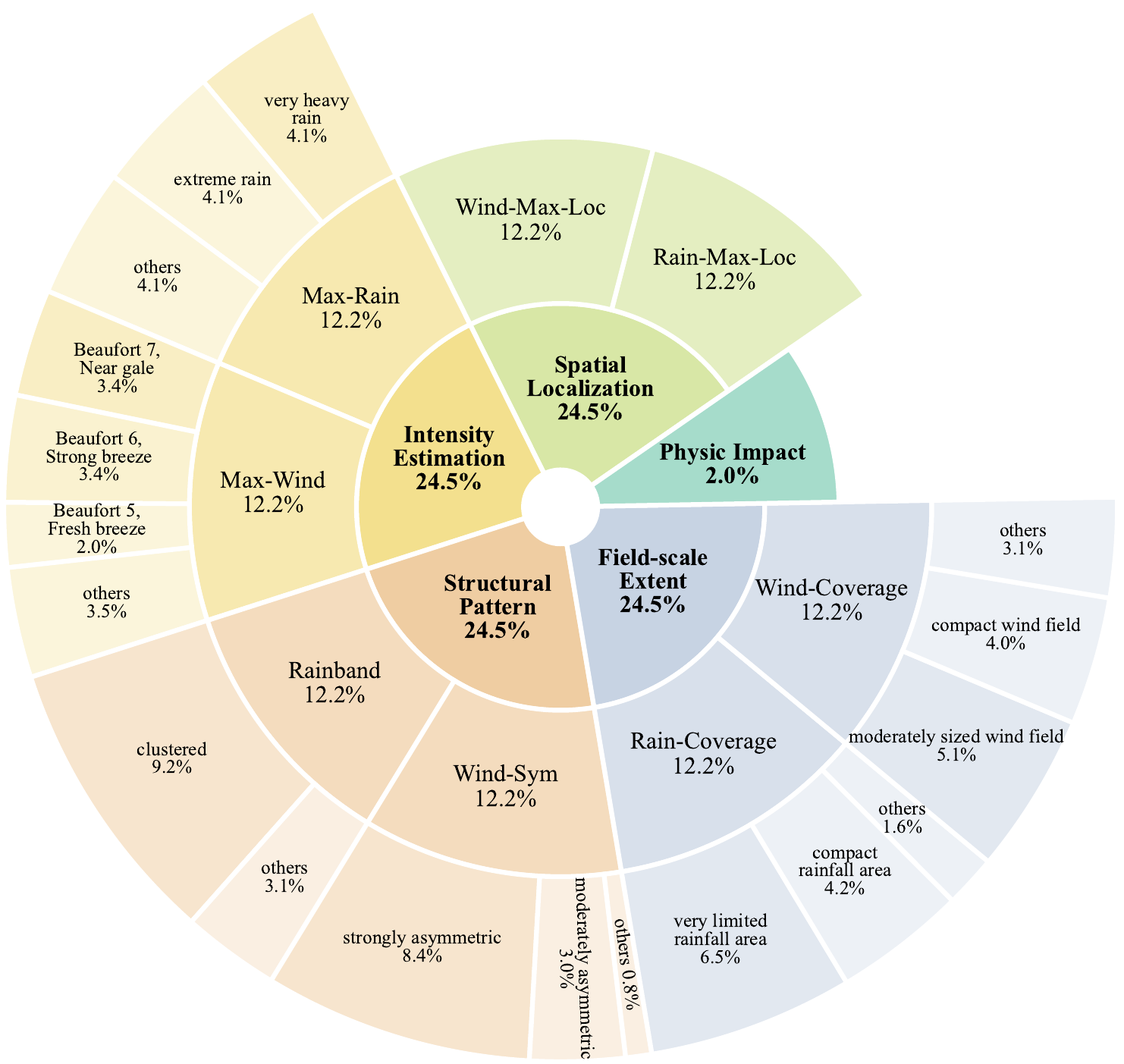}
    \caption{Task and label distribution of OmniClimate-TC. Label classes below 2\% are grouped as \textit{Others} for visualization.}
    \label{fig:data_dis}
\end{figure}

\textbf{Dataset statistics.}
Our proposed OmniClimate-TC comprises 4,978 high-resolution wind and precipitation heatmaps derived from 2,489 tropical cyclone events, covering $\sim$130 countries and 46 oceanic regions (Fig.~\ref{fig:climate_tc_map}).
From these inputs, we construct 243,890 instruction-tuning examples spanning five task categories, with perceptual tasks evenly distributed (24.5\% each) and Physical Impact Analysis accounting for 2\% (Fig.~\ref{fig:data_dis}).
Data are split chronologically into training (1979--2015), validation (2016--2019), and test (2020--2025) sets, with proportions approximately following a 7:1:2 ratio.


\subsection{ClimateTCX: Two-Stage Instruction Tuning}

Each example consists of two hazard heatmaps (wind and precipitation) paired with a textual instruction.
We consider eight \emph{perceptual} tasks spanning spatial localization, intensity, structure, and field-scale extent, together with one \emph{compositional} reasoning task for physical impact analysis.

\textbf{Stage-1: Vision Perceptual Alignment.}
In Stage-1, we fine-tune the model on the eight perceptual tasks using supervised instruction tuning.
Let $\mathcal{D}_1=\{(x_i,y_i,t_i)\}$ denote the training set, where $x_i$ is the multimodal input and $t_i$ the task label.
We optimize the autoregressive objective:
\begin{equation}
\small
\mathcal{L}_{\mathrm{SFT}}(\theta)
= - \mathbb{E}_{(x,y,t)\sim \mathcal{D}_1}
\sum_{j} \log p_{\theta}(y_j \mid x, y_{<j}),
\end{equation}
using task-balanced sampling to avoid dominance by high-frequency tasks.
This stage establishes visually grounded perceptual primitives for interpreting cyclone structure.

\textbf{Stage-2: Mixed Training with Impact Reasoning.}
Stage-2 continues training from the Stage-1 checkpoint and introduces the physical impact
analysis task $t_{\mathrm{imp}}$, which requires synthesizing wind and precipitation cues
into a mechanism-focused hazard summary.
Training is performed on a mixed dataset $\mathcal{D}_2$ with a task mixture:
\begin{equation}
\small
P(t)=
\begin{cases}
\rho, & t = t_{\mathrm{imp}},\\[3pt]
\dfrac{1-\rho}{8}, & t \in \mathcal{T}_{\mathrm{perc}},
\end{cases}
\end{equation}
which is implemented via weighted sampling $w_i \propto P(t_i)/N(t_i)$.
Stage-2 mixes impact and perceptual supervision to promote compositional transfer from visually grounded cues to physical hazard interpretation without catastrophic forgetting.

\begin{table*}[t]
\caption{
Model performance on OmniClimate-TC.
For spatial localization, we report multi-scale scores (EMS$_c$, EMS$_a$; higher is better, values closer to 0 indicate more accurate localization).
For other tasks, we report Accuracy and Macro-F1 (\%).
Overall Acc/F1 average the six non-localization tasks, and Overall EMS$_t$ averages Rain-Max-Loc and Wind-Max-Loc.
Within each model category (open-source, closed-source, ours), the best and second-best results are highlighted in \textbf{bold} and \underline{underlined}.
}
\label{tab:tc_visual_model_perf_ms}
\centering
\scriptsize
\setlength{\tabcolsep}{2.6pt}
\renewcommand{\arraystretch}{1.05}
\resizebox{0.95\textwidth}{!}{%
\begin{tabular}{@{}ll*{19}{c}@{}}
\toprule
 &  & \multicolumn{4}{c}{\textbf{Spatial Localization}}
 & \multicolumn{4}{c}{\textbf{Intensity Estimation}}
 & \multicolumn{4}{c}{\textbf{Structural Pattern}}
 & \multicolumn{4}{c}{\textbf{Field-scale Extent}}
 & \multicolumn{3}{c}{\textbf{Overall}} \\
\cmidrule(lr){3-6}\cmidrule(lr){7-10}\cmidrule(lr){11-14}\cmidrule(lr){15-18}\cmidrule(lr){19-21}
\textbf{Category} & \textbf{Model}
 & \multicolumn{2}{c}{Rain-Max-Loc} & \multicolumn{2}{c}{Wind-Max-Loc}
 & \multicolumn{2}{c}{Max-Rain} & \multicolumn{2}{c}{Max-Wind}
 & \multicolumn{2}{c}{Rainband} & \multicolumn{2}{c}{Wind-Sym}
 & \multicolumn{2}{c}{Rain-Cov} & \multicolumn{2}{c}{Wind-Cov}
 & Acc$\uparrow$ & F1$\uparrow$ & EMS$_{t}\uparrow$ \\
 & 
 & EMS$_{c}\uparrow$ & EMS$_{a}\uparrow$
 & EMS$_{c}\uparrow$ & EMS$_{a}\uparrow$
 & Acc$\uparrow$ & F1$\uparrow$ & Acc$\uparrow$ & F1$\uparrow$
 & Acc$\uparrow$ & F1$\uparrow$ & Acc$\uparrow$ & F1$\uparrow$
 & Acc$\uparrow$ & F1$\uparrow$ & Acc$\uparrow$ & F1$\uparrow$
 &  &  &  \\
\midrule

\multirow{10}{*}{\textbf{Open-source}}
& InternVL3.5-14B
& -0.431 & -0.986 & \best{-0.301} & -0.991
& 31.0 & 10.2 & \second{19.8} & \second{11.5}
& 35.7 & 18.1 & \second{53.4} & 21.2
& 22.1 & 12.1 & 18.4 & 10.6
& 30.1 & 14.0 & -0.773 \\

& InternVL3.5-38B
& -0.618 & -0.986 & -0.543 & -0.979
& 28.9 & 13.8 & \best{20.7} & \best{13.9}
& 35.9 & 14.1 & 46.2 & 20.4
& \best{30.8} & 16.4 & 20.0 & 9.9
& 30.4 & \second{14.8} & -0.837 \\

& LLaVA-v1.6-13B
& -0.646 & -1.000 & -0.795 & -1.000
& 30.4 & 10.3 & 0.0 & 0.0
& 20.5 & 9.4 & \best{59.0} & \second{22.3}
& 2.9 & 2.8 & 16.6 & 8.5
& 21.5 & 8.9 & -0.907 \\

& LLaVA-v1.6-34B
& -0.756 & -0.992 & -0.871 & -0.999
& 7.8 & 3.1 & 0.0 & 0.0
& 20.1 & 10.6 & 16.2 & 16.8
& 20.7 & 12.8 & \best{28.9} & \second{15.4}
& 15.6 & 9.8 & -0.932 \\

& Qwen2.5-VL-7B
& -0.361 & -0.980 & -0.392 & -0.974
& 33.2 & 8.3 & 1.2 & 1.1
& \second{55.6} & 15.9 & 23.9 & 9.8
& 2.7 & 2.0 & 19.1 & 9.1
& 22.6 & 7.7 & -0.762 \\

& Qwen2.5-VL-32B
& \second{-0.239} & \best{-0.946} & -0.346 & \best{-0.959}
& \best{37.1} & \second{17.8} & 0.0 & 0.0
& 50.6 & 22.8 & 25.8 & 15.8
& 15.5 & 15.3 & \second{28.8} & \best{15.8}
& 26.3 & 14.6 & \best{-0.700} \\

& Qwen2.5-VL-72B
& \best{-0.235} & -0.973 & -0.497 & \best{-0.959}
& 33.6 & 10.8 & 0.7 & 1.5
& \best{65.1} & \second{23.4} & 38.2 & 19.8
& 27.5 & 19.6 & 25.1 & 13.8
& \best{31.7} & \second{14.8} & -0.745 \\

& Qwen3-VL-8B
& -0.308 & -0.976 & -0.413 & -0.968
& 33.4 & 11.6 & 0.1 & 0.1
& 37.5 & 18.0 & 49.1 & \best{25.6}
& 18.6 & 11.5 & 19.7 & 13.1
& 26.4 & 13.3 & -0.750 \\

& Qwen3-VL-32B
& -0.242 & -0.967 & \second{-0.327} & -0.964
& 36.2 & \second{17.8} & 0.3 & 0.7
& 32.2 & 17.1 & 53.1 & 18.7
& 20.9 & 12.7 & 23.5 & 11.9
& 27.7 & 13.2 & \second{-0.717} \\

& Qwen3-VL-235B
& -0.284 & \second{-0.963} & -0.420 & \second{-0.963}
& \second{36.6} & \best{21.8} & 5.3 & 6.3
& 51.4 & \best{23.8} & 40.9 & 20.8
& \second{30.3} & \best{19.9} & 18.5 & 7.2
& \second{30.5} & \best{16.6} & -0.742 \\

\midrule
\multirow{8}{*}{\textbf{Closed-source}}
& GPT-4o-mini
& -0.329 & -0.967 & -0.380 & -0.970
& 34.7 & 15.9 & 7.9 & 4.8
& 64.8 & 23.6 & 59.0 & 27.3
& 6.5 & 6.3 & 18.4 & 10.3
& 31.9 & 14.7 & -0.759 \\

& GPT-4o
& \second{-0.175} & -0.960 & -0.552 & -0.977
& 37.8 & 15.6 & 2.3 & 1.6
& \second{68.8} & 21.2 & 23.5 & 13.0
& 29.6 & 21.4 & 15.6 & 7.6
& 29.6 & 13.4 & -0.746 \\

& GPT-5-mini
& -0.268 & -0.937 & -0.487 & -0.963
& \second{45.0} & 16.0 & \second{29.1} & \second{12.1}
& 68.5 & \best{24.8} & 62.7 & \second{35.2}
& 36.1 & 23.4 & \second{19.1} & \best{17.3}
& \second{43.4} & \best{21.5} & -0.724 \\

& GPT-5.1
& \best{-0.170} & \second{-0.921} & \best{-0.361} & \second{-0.953}
& 42.7 & \best{19.9} & 5.4 & 5.2
& 64.3 & \second{23.7} & 41.3 & 28.6
& \second{38.9} & \best{30.3} & 16.8 & \second{12.2}
& 34.9 & \second{20.0} & \second{-0.678} \\

& Gemini-2.0-Flash
& -0.487 & -0.988 & \second{-0.375} & -0.979
& 31.8 & 16.0 & 9.4 & 5.8
& 64.7 & 18.1 & 36.1 & 20.8
& 28.0 & 18.3 & 14.0 & 5.7
& 30.7 & 14.1 & -0.791 \\

& Gemini-2.5-Flash
& -0.287 & -0.944 & -0.594 & -0.981
& 44.5 & 16.3 & 7.9 & 5.0
& 66.7 & 18.8 & \best{68.1} & \best{36.8}
& 36.6 & 19.9 & 10.0 & 8.6
& 39.0 & 17.6 & -0.766 \\

& Gemini-Lite-Latest
& -0.361 & -0.965 & -0.403 & -0.979
& 31.9 & 14.7 & 4.7 & 3.8
& 68.3 & 18.3 & 28.4 & 11.1
& 37.1 & 23.2 & \best{19.6} & 11.8
& 31.7 & 13.8 & -0.759 \\

& Gemini-3-Flash
& -0.226 & \best{-0.846} & -0.580 & \best{-0.939}
& \best{45.7} & \second{17.9} & \best{34.3} & \best{16.2}
& \best{69.5} & 19.4 & \second{65.0} & 21.2
& \best{42.7} & \second{26.3} & 15.9 & 10.2
& \best{45.5} & 18.5 & \best{-0.660} \\

\midrule
\multirow{3}{*}{\textbf{ClimateTCX (Ours)}}
& LLaVA-v1.6-34B
& -0.389 & -0.932 & -0.543 & -0.942
& 37.8 & 23.9 & 13.1 & 12.7
& 76.2 & 27.3 & \second{82.5} & 54.3
& \second{72.0} & 51.2 & 66.4 & \second{33.3}
& 58.0 & 33.8 & -0.738 \\

& Qwen2.5-VL-32B
& \best{-0.108} & \second{-0.714} & \best{-0.075} & \second{-0.717}
& \second{57.3} & \second{42.7} & \best{51.0} & \second{31.0}
& \second{79.7} & \second{24.9} & 81.5 & \second{67.0}
& \best{74.2} & \best{60.5} & \best{69.2} & \best{39.3}
& \best{68.9} & \second{44.2} & \second{-0.558} \\

& Qwen3-VL-32B
& \second{-0.119} & \best{-0.713} & \second{-0.122} & \best{-0.697}
& \best{58.3} & \best{43.4} & \second{47.7} & \best{31.1}
& \best{79.8} & \best{32.6} & \best{82.5} & \best{70.8}
& 69.7 & \second{60.3} & \second{66.4} & 33.1
& \second{67.4} & \best{45.2} & \best{-0.554} \\

\bottomrule
\end{tabular}%
}
\end{table*}

\begin{table}[t]
\centering
\small
\setlength{\tabcolsep}{4.5pt}
\caption{\textbf{Physical impact analysis evaluation.} ClimateTCX variants are based on Qwen3-VL-32B.
We group metrics into physics-aware rule-based, lexical text similarity, and LLM-as-judge. Higher is better ($\uparrow$).}
\label{tab:impact_eval_grouped}
\resizebox{0.45\textwidth}{!}{
\begin{tabular}{l cc cc c}
\toprule
& \multicolumn{2}{c}{\textbf{Physics-aware}}
& \multicolumn{2}{c}{\textbf{Text Similarity}}
& \multicolumn{1}{c}{\textbf{LLM-as-Judge}} \\
\cmidrule(lr){2-3} \cmidrule(lr){4-5} \cmidrule(lr){6-6}
\textbf{Model}
& \textbf{HSF}$\uparrow$
& \textbf{HAA}$\uparrow$
& \textbf{BLEU}$\uparrow$
& \textbf{ROUGE}$\uparrow$
& \textbf{Gemini Score}$\uparrow$ \\
\midrule

\multicolumn{6}{l}{\textbf{Open-source models}}\\
\midrule
Qwen3-VL-32B      & 0.1612 & 0.2207 & 0.0268 & 0.1670 & 0.4457 \\
Qwen2.5-VL-32B   & 0.1440 & 0.1808 & 0.0251 & 0.1361 & 0.4268 \\
LLaVA-v1.6-34B   & 0.1627 & 0.2605 & 0.0232 & 0.1489 & 0.5221 \\

\midrule
\multicolumn{6}{l}{\textbf{Closed-source models}}\\
\midrule
GPT-5.1              & 0.2097 & 0.3873 & 0.0337 & 0.1799 & 0.6876 \\
Gemini-3-Flash       & 0.1886 & 0.3075 & 0.0186 & 0.1044 & 0.6530 \\
GPT-4o               & 0.2089 & 0.4366 & 0.0196 & 0.1092 & 0.6419 \\
GPT-5-mini         & 0.1964 & 0.3568 & 0.0239 & 0.1325 & 0.5895 \\
Gemini-Lite-Latest   & 0.1338 & 0.2606 & 0.0253 & 0.1344 & 0.5719 \\
Gemini-2.5-Flash     & 0.1017 & 0.1221 & 0.0144 & 0.0908 & 0.4114 \\

\midrule
ClimateTCX\_MIX
& \textbf{0.2496}
& \textbf{0.5000}
& \underline{0.1326}
& \underline{0.3183}
& \underline{0.7514} \\

ClimateTCX\_VPA
& 0.1573
& 0.3779
& 0.0236
& 0.1202
& 0.6425 \\

ClimateTCX\_MT
& \underline{0.2449}
& \underline{0.4906}
& \textbf{0.1347}
& \textbf{0.3233}
& \textbf{0.7599} \\
\bottomrule
\end{tabular}
}
\end{table}

\begin{table*}[t]
\centering
\caption{\textbf{Ablation study on ClimateTCX (Qwen3-VL-32B).}
For Spatial Localization, EMS$_c$ and EMS$_a$ are averaged over Rain-Max-Loc and Wind-Max-Loc.
For other categories, Accuracy and Macro-F1 are averaged over the two tasks within each category.
Best and second-best results across ablation settings are shown in \best{bold} and \second{underline}, respectively.}
\label{tab:tc_visual_ablation_compact}
\small
\setlength{\tabcolsep}{5pt}
\renewcommand{\arraystretch}{1.15}
\resizebox{0.9\textwidth}{!}{%
\begin{tabular}{lcc|cc|cc|cc|cccc}
\toprule
\textbf{Setting}
& \multicolumn{2}{c|}{\textbf{Spatial}} 
& \multicolumn{2}{c|}{\textbf{Intensity}}
& \multicolumn{2}{c|}{\textbf{Structural}}
& \multicolumn{2}{c|}{\textbf{Extent}}
& \multicolumn{4}{c}{\textbf{Impact Analysis}}\\
\cmidrule(lr){2-3}\cmidrule(lr){4-5}\cmidrule(lr){6-7}\cmidrule(lr){8-9}\cmidrule(lr){10-13}
& EMS$_c \uparrow$ & EMS$_a \uparrow$
& Acc$\uparrow$ & F1$\uparrow$
& Acc$\uparrow$ & F1$\uparrow$
& Acc$\uparrow$ & F1$\uparrow$ 
& HSF$\uparrow$ & HAA$\uparrow$ & ROUGE$\uparrow$ & Gemini Score$\uparrow$ \\
\midrule
Full (no ablation)
& \second{-0.121} & \best{-0.705}
& \best{53.0} & \best{37.3}
& \best{81.2} & 51.7
& 68.1 & \best{46.7}
& \best{0.2449} & \best{0.4906} & \best{0.3233} & \best{0.7599} \\

w/o Extent
& \best{-0.118} & \second{-0.713}
& \second{52.6} & \second{37.1}
& \second{80.7} & \second{53.7}
& 0.1 & 0.3
& 0.2351 & 0.4847 & 0.3014 & 0.7490 \\

w/o Intensity
& -0.128 & -0.716
& 11.9 & 7.8
& 80.9 & 53.5
& \best{69.3} & \second{46.4}
& 0.2300 & 0.4742 & 0.3138 & 0.7466 \\

w/o Spatial
& -0.368 & -0.767
& 52.3 & 36.5
& 80.0 & \best{54.1}
& \second{68.6} & 46.1
& \second{0.2404} & \second{0.4870} & 0.3152 & \second{0.7506} \\

w/o Structural
& -0.144 & \second{-0.707}
& 52.3 & \second{37.1}
& 7.1 & 7.3
& 67.6 & 45.5
& 0.2324 & 0.4707 & \second{0.3173} & 0.7452 \\
\bottomrule
\end{tabular}
}
\end{table*}

\section{Experiment}
\textbf{Experimental Setup}.
We fine-tune three representative VLMs:
LLaVA-v1.6-34B~\cite{liu2023improved},
Qwen2.5-VL-32B-Instruct~\cite{qwen2.5}, and
Qwen3-VL-32B-Instruct~\cite{qwen3technicalreport}.
We further evaluate a range of open- and closed-source VLMs on OmniClimate-TC,
including models from the Qwen, LLaVA, InternVL, ChatGPT, and Gemini families
\cite{wang2025internvl3_5,achiam2023gpt,team2023gemini}.
All compared models are evaluated on the same benchmark tasks; only ClimateTCX is trained on OmniClimate-TC.
Implementation details are in Appendix.

\subsection{Evaluation Metrics}

We evaluate the eight perception tasks using task-aligned metrics that reflect their physical and semantic requirements.

\textbf{Classification-style tasks.}
Three task categories (Intensity Estimation, Structural Pattern, and Field-scale Extent) are formulated as closed-set classification problems.
We report \emph{Accuracy} and \emph{Macro-F1} to measure exact label agreement.

\textbf{Spatial localization tasks.}
Spatial Localization requires identifying the geographic location of hazard extrema.
Predictions are evaluated at multiple administrative scales using an Element Match Score (EMS)~\cite{chen2025climateiqa}.
We report EMS at the country level (EMS$_c$), admin-1 level (EMS$_a$), and their aggregate (EMS$_t$).



\textbf{Physics-aware rule-based metrics.}
We evaluate physical impact analysis using complementary metrics that assess faithfulness and hazard attribution.
\textbf{Hazard Severity Faithfulness (HSF)} measures whether stated wind and rainfall severities are supported by the underlying PAVA-derived physical attributes, penalizing hallucinated or exaggerated claims~\cite{maynez2020faithfulness}.
\textbf{Hazard Attribution Accuracy (HAA)} evaluates whether the dominant hazard mechanism is correctly identified, consistent with IPCC definitions of multi-hazard impacts~\cite{change2001climate}.
Both metrics rely on a physics-grounded hazard-impact lexicon curated from NOAA/NWS, WMO, and IPCC (see Appendix).

\textbf{Lexical similarity metrics.}
We additionally report BLEU~\cite{papineni2002bleu} and ROUGE~\cite{lin2004rouge} to quantify surface-level similarity with reference summaries. \rev{Because the references are generated under fixed GPT-5.1 prompt constraints, BLEU and ROUGE are best interpreted as measures of agreement with a controlled reference style rather than direct evidence of physical validity.}

\textbf{LLM-as-judge evaluation.}
We employ an LLM-based judge (Gemini-3-Pro) to assess higher-level semantic properties, including attribute consistency, dominant hazard correctness, mechanism support, and overall completeness. 
\rev{Here the judge serves as a constrained semantic verifier conditioned only on PAVA-derived attributes. The expert audit reported in Fig.~\ref{fig:expert_eval} and Tab.~\ref{tab:expert_factual_audit} therefore acts as the primary external check on target quality, while lexical and judge-based scores are secondary diagnostics.} Details are provided in the Supplementary material.

\subsection{Main results.}

\textbf{Perception tasks.} \rev{Tab.~\ref{tab:tc_visual_model_perf_ms} shows that general-purpose VLMs remain poorly matched to tropical-cyclone hazard fields. Open-source models display substantial variability across tasks: some capture salient morphology, but most fail on physically calibrated quantities such as peak-wind intensity and field extent, often approaching chance-level performance. These patterns suggest that visual saliency alone does not yield stable understanding of magnitude or spatial footprint. Closed-source models improve aggregate accuracy and are often effective on mesoscale pattern recognition, yet they continue to exhibit systematic weakness in intensity grounding and extent reasoning, consistent with a bias toward shape-level cues rather than field-consistent physical interpretation.}
\rev{In contrast, \textbf{ClimateTCX} achieves the strongest and most balanced results across localization, intensity, structure, and extent, supporting PAVA-aligned representation design and supervision as an effective interface for reasoning over hazard fields.}

\begin{figure}[t]
\centering
\includegraphics[width=\columnwidth]{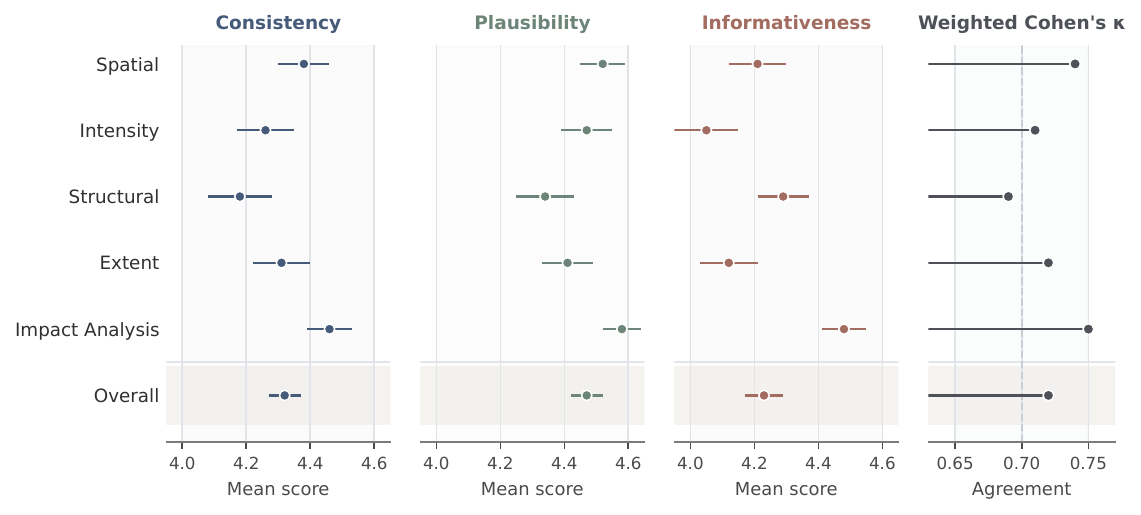}
\caption{
Expert evaluation of OmniClimate-TC across task categories.
The first three panels report consistency, plausibility, and informativeness on a 5-point Likert scale, and the fourth reports weighted Cohen's $\kappa$; points show means with 95\% bootstrap confidence intervals.
}
\label{fig:expert_eval}
\end{figure}

\textbf{Physical impact analysis.}
\rev{Tab.~\ref{tab:impact_eval_grouped} underscores the difficulty of converting visual hazard cues into physically faithful impact summaries. Open-source models perform poorly on HSF and HAA, indicating frequent overstatement of hazard severity or incorrect attribution of the dominant mechanism despite reasonable fluency. Closed-source models improve both faithfulness and attribution, but a substantial gap remains between lexical similarity and physics-aware correctness, showing that fluent text is not, by itself, evidence of physical consistency. ClimateTCX performs best across metric groups. Stage-1 perceptual training alone (ClimateTCX\_VPA) yields clear gains over generic baselines, indicating that perceptual grounding transfers to impact reasoning. Adding impact supervision (ClimateTCX\_MIX) improves attribution and textual quality, while the proposed mixed strategy (ClimateTCX\_MT) provides the strongest overall balance among faithfulness, hazard attribution, and judge-based semantic consistency. This interpretation is consistent with the expert verification reported in Fig.~\ref{fig:expert_eval} and Tab.~\ref{tab:expert_factual_audit}, which provide the primary external check on benchmark fidelity.}

\rev{In conclusion, these results suggest that accurate physical impact analysis depends more on perceptual grounding and task composition than on model scale or linguistic fluency alone.}

\subsection{Ablation analysis.}
\rev{Tab.~\ref{tab:tc_visual_ablation_compact} shows that each perceptual abstraction contributes measurably to both task-specific prediction and downstream impact reasoning. Removing any single component degrades not only the corresponding perceptual task, but also the quality of physical impact analysis, indicating that these abstractions function as a coupled representation rather than as independent auxiliaries. Extent removal produces the sharpest decline in extent prediction and the largest downstream losses in ROUGE and Gemini Score, underscoring the importance of hazard footprint in judging impact scale and spatial reach. Removing intensity substantially weakens magnitude recognition and yields consistent losses in HSF and HAA, showing that calibrated severity cues are central to faithful hazard interpretation. Structural ablation primarily impairs mesoscale pattern recognition and propagates to impact reasoning, consistent with the role of organization and asymmetry in determining dominant hazard mechanisms. Spatial ablation mainly affects localization metrics and has comparatively smaller impact on abstract impact synthesis, suggesting that for this task, geolocation is less consequential than magnitude, structure, and extent. Overall, the full model yields the strongest balance across metrics, indicating that robust physical impact understanding depends on the joint availability of complementary perceptual abstractions.}

\subsection{Expert evaluation and Dataset Reliability}
\rev{We assess benchmark reliability by sampling 40 TC cases and reviewing the nine associated tasks per case, yielding 360 expert-reviewed instances. Two tropical-meteorology experts independently rate visual–physical consistency, physical plausibility, and informativeness on a 5-point Likert scale. Fig.~\ref{fig:expert_eval} shows consistently strong scores across categories, with all three criteria above 4.0 and narrow confidence intervals.}

\rev{
Weighted Cohen’s $\kappa$ ranges from 0.69 to 0.75 across categories, with an overall $\kappa$ of 0.72. Overall scores reach 4.32 for consistency, 4.47 for plausibility, and 4.23 for informativeness, supporting the reliability of OmniClimate-TC.
}
\begin{table}[t]
\centering
\caption{\rev{Expert verification of generated perceptual labels and impact-summary acceptability on 40 sampled TC cases (360 reviewed instances).}}
\label{tab:expert_factual_audit}
\small
\renewcommand{\arraystretch}{1.08}
\begin{tabular*}{\columnwidth}{@{\extracolsep{\fill}}lcc}
\toprule
\rev{Task} & \rev{Correct/Total} & \rev{Rate (\%)} \\
\midrule
\rev{Spatial} & \rev{68/80} & \rev{85.0} \\
\rev{Intensity} & \rev{69/80} & \rev{86.3} \\
\rev{Structural} & \rev{73/80} & \rev{91.3} \\
\rev{Extent} & \rev{67/80} & \rev{83.8} \\
\midrule
\rev{Perceptual overall} & \rev{277/320} & \rev{86.6} \\
\rev{Impact reasoning} & \rev{33/40} & \rev{82.5} \\
\bottomrule
\end{tabular*}
\end{table}
\rev{Tab.~\ref{tab:expert_factual_audit} further audits whether the generated targets remain faithful to the underlying physical labels. Perceptual annotations achieve 86.6\% expert-verified accuracy overall. 
Impact summaries achieve 82.5\% expert acceptability. Fig.~\ref{fig:expert_eval} and Tab.~\ref{tab:expert_factual_audit} support the reliability of the annotations and generated targets for benchmark construction and instruction tuning.}


\begin{figure}[t]
\centering
\includegraphics[width=0.9\columnwidth]{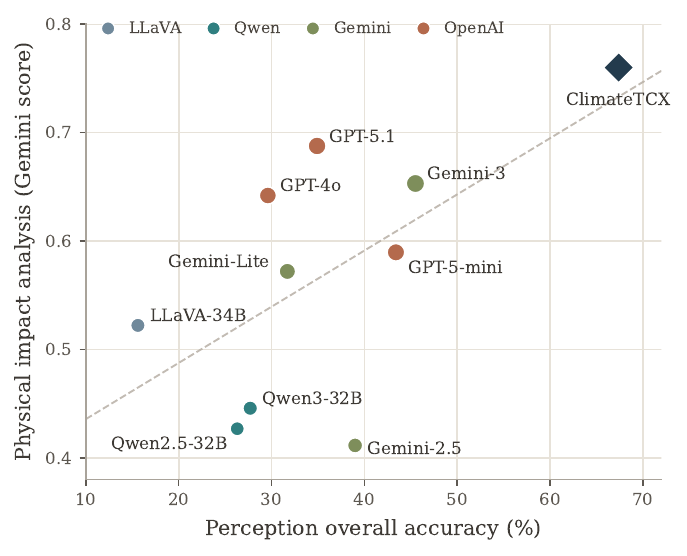}
\caption{\rev{Perception--reasoning relation across representative open- and closed-source models.}}
\label{fig:perception_reasoning_relation}
\end{figure}

\subsection{Perception--reasoning relation}
\rev{Fig.~\ref{fig:perception_reasoning_relation} shows the relation between perceptual grounding and downstream impact reasoning across model families. Models with stronger event-level perception generally achieve better impact-analysis performance. This descriptive association supports our premise that reliable impact reasoning benefits from accurate perceptual grounding, without implying causality.}

\rev{The figure also shows that this relation is not explained by model family or scale alone. Closed-source models generally occupy a stronger regime than open-source baselines, but their gains remain uneven: some attain moderate impact-analysis scores despite only modest perceptual accuracy, whereas others improve perception without a commensurate gain in reasoning quality. ClimateTCX lies in the upper-right corner and is separated from the rest of the model set on both axes, indicating that the proposed training strategy improves not only perceptual recognition, but also the conversion of perceptual evidence into coherent physical impact judgments. In this sense, the figure complements Tabs.~\ref{tab:tc_visual_model_perf_ms} and~\ref{tab:impact_eval_grouped} by showing that the strongest impact-analysis behavior is attained when perceptual grounding and reasoning are improved jointly rather than in isolation.}

\section{Conclusion}

\rev{This study frames multimodal extreme-weather understanding as a representation problem. We introduce PAVA, a physics-grounded interface that makes meaningful structure in continuous geophysical fields accessible to VLMs. PAVA is studied primarily as a supervision and evaluation interface rather than as an auxiliary inference-time modality. Built on this interface, OmniClimate-TC and ClimateTCX show that PAVA-aligned supervision improves both perceptual prediction and impact reasoning over tropical-cyclone hazard fields. More broadly, our results highlight representation design, beyond model scale, as a key factor in multimedia reasoning over scientific fields. Our study is limited to tropical cyclones and the current PAVA-aligned annotation pipeline; disentangling PAVA from broader domain adaptation will require matched non-PAVA controls. Future work will extend this interface-design recipe to other hazards and annotation schemes.}

\section*{Acknowledgments}
GM acknowledges MOE Tier 2 grant no. T2EP20124-0043: ARTIFICIAL INTELLIGENCE + DATA ASSIMILATION (AIDA) FOR WEATHER AND CLIMATE MODELS.
\bibliographystyle{ACM-Reference-Format}
\bibliography{references}

\end{document}